# Manipulating Steady Heat Conduction by *Sensu*-shaped Thermal Metamaterials

Tiancheng Han[1*], Xue Bai[1,2,3*], Dan Liu[1,2,3*], Dongliang Gao[1], Baowen Li[2,3,4], John T. L. Thong[1,3], and Cheng-Wei Qiu[1,3]

[1]Department of Electrical and Computer Engineering, National University of Singapore, 4 Engineering Drive 3, Republic of Singapore.    *Equal Contribution

[2]Department of Physics and Centre for Computational Science and Engineering, National University of Singapore, Singapore 117546, Republic of Singapore

[3]NUS Graduate School for Integrative Sciences and Engineering, National University of Singapore, Kent Ridge 119620, Republic of Singapore.

[4]Center for Phononics and Thermal Energy Science, School of Physics Science and Engineering, Tongji University, 200092, Shanghai, China.

The ability to design the control of heat flow has innumerable benefits in the design of electronic systems such as thermoelectric energy harvesters, solid-state lighting, and thermal imagers, where the thermal design plays a key role in performance and device reliability. However, to realize one advanced control function of thermal flux, one needs to design one sophisticated, multilayered and inhomogeneous thermal structure with different composition/shape at different regions of one device. In this work, we employ one identical sensu-unit with facile natural composition to experimentally realize a new class of thermal metamaterials for controlling thermal conduction (e.g., thermal concentrator, focusing/resolving, uniform heating), only resorting to positioning and locating the same unit element of sensu-shape structure. The thermal metamaterial unit and the proper arrangement of multiple identical units are capable of transferring, redistributing and managing thermal energy in a versatile fashion. It is also shown that our sensu-shape unit elements can be used in manipulating dc currents without any change in the layout for the thermal counterpart. The proposed scheme can also be applied to control dc electric currents and dc magnetic fields that governed by Laplace equation. These could markedly enhance the capabilities in thermal sensing, thermal imaging, thermal-energy storage, thermal packaging, thermal therapy, and more domains beyond.

Researchers have been pursuing effective methodologies to control thermal flux for multifarious applications[1-13]. The manipulation of heat flow is essential in technology development in many areas such as thermoelectricty[1], fuel cells[2], thermal barrier coatings[3], solar cells[4], electronics and optoelectronics[5], and low thermal conductivity materials[6]. In addition, the ability to precisely control heat flow can potentially lead to the development of thermal analogues of electrical circuit components[7], such as thermal diodes[8-10], thermal transistors[11], and thermal memory[12]. More recently, thermo-crystals were theoretically proposed for thermal management, guiding thermal wave just as photonic crystals guide light[13].

While the conduction of heat has been known for a long time, the advance in the control of heat flow has been very slow[7-16]. Most recently, by using a multilayered composite approach, cloaking, concentration, and reversal of heat flux were experimentally demonstrated in thick composites[17] and planar structures[18]. Considering that the heat conduction equation is form invariant under coordinate transformations, thermal cloaks were realized with inhomogeneous anisotropic thermal conductivities[19, 20]. On the basis of exact methodology (directly solving heat conduction equation), bilayer thermal cloaks made of bulk isotropic materials have been reported[21, 22]. Though significant progress has been made toward the manipulation of heat flow, different functionalities have to rely on different structures[14-23]. This motivates us to explore a general class of thermal metamaterial units, with which a set of interesting functionalities in advanced control of heat conduction can be experimentally demonstrated by positioning and combining identical unit.

In this paper, we introduce a new class of thermal metamaterials composed of two regular materials in a simplified planar geometry. The thermal metamaterial unit, designed with transformation optics[24, 25], is capable of manipulating thermal energy and heat flux in a versatile variety of fashions by positioning and combining identical unit *sensu*-elements (which is a traditional folding hand-fan in Japan). We experimentally demonstrate that the combination of thermal metamaterial units exhibits novel properties such as the creation of a uniform heating region, heat flux focusing, and heat flux concentration. On one hand, these proposed metamaterials are fabricated using regular materials and can hence be easily accessed and followed. On the other hand, these novel properties are remarkably robust to the geometrical sizing of the proposed metamaterials without having to change the material compositions.

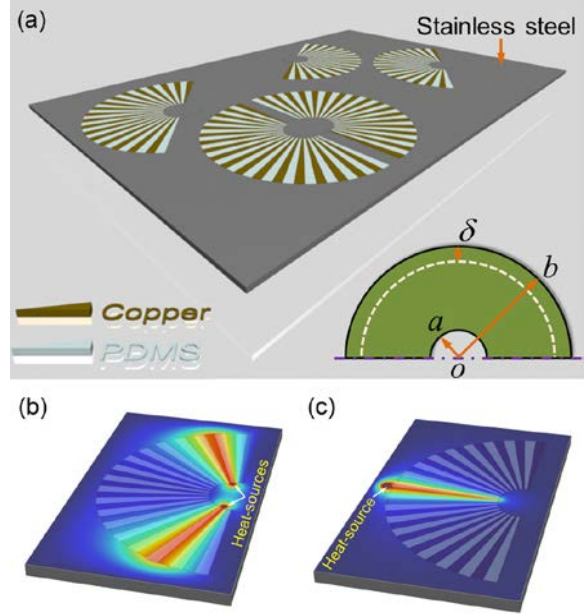

**FIG. 1** (a) Schematic of a random cluster of sensu-shaped thermal metamaterial units made of two regular bulk materials. The inset schematically shows the transformation principle of thermal metamaterial unit. (b) Two heat-sources placed near the inner boundary of a thermal metamaterial unit, analogous to the performance of an optical hyperlens. (c) A heat-source placed near the outer boundary of a thermal metamaterial unit, demonstrating harvesting property.

Fig. 1(a) shows the schematic of a random cluster of *sensu*-shaped thermal metamaterial (SSTM) units made of two regular bulk materials. The design of a SSTM unit is based on transformation optics[24, 25], which is schematically illustrated in the inset of Fig. 1(a). The semi-annular region ($b-\delta \leq r \leq b$) in virtual space is extruded to a larger region ($a \leq r \leq b$) in real space, generating anisotropic thermal conductivity ($\kappa_r \to \infty$ and $\kappa_\theta \to 0$) in the region ($a \leq r \leq b$) when $\delta \to 0$. It is rational to set $\kappa_r = \kappa_b 2^n$ and $\kappa_\theta = \kappa_b 2^{-n}$ so long as $n$ is large enough, where $\kappa_b$ is the thermal conductivity of the background. On the basis of effective medium theory (EMT), an anisotropic material may be practically realized by alternately stacking two materials A and B in the azimuthal direction. Assuming that the background is stainless steel, we choose copper and polydimethylsiloxane (PDMS) as materials A and B, respectively. Thus we can obtain the SSTM unit composed of 18 copper wedges and 18 PDMS wedges, shown in Fig. 1(a). The conductivities of copper, PDMS, and stainless steel are $\kappa_{Cu} = 394$ W/Km,

$\kappa_{PDMS} = 0.15$ W/Km, and $\kappa_b = 16$ W/Km, respectively. We choose *a*=1.2 cm and *b*=6 cm for both simulations and experiments throughout.

To understand the functionalities of the SSTM unit, numerical simulations based on finite element modelling are demonstrated in Figs. 1(b) and 1(c). In the simulation setup, the four boundaries of the stainless steel (background) are set as fixed temperature at $0\ ^\circ C$, which ensures that the heat is mainly conducted rather than convected. When two point heat-sources are placed near the inner boundary of a SSTM unit, the simulated thermal profile is shown in Fig. 1(b). It is clearly seen that the thermal fields of the two heat-sources are almost perfectly transferred from inner-boundary to outer-boundary, analogous to the performance of an optical or acoustic hyperlens[26-29]. In contrast to the conceptual thermocrystal superlens for near-field imaging[13], our scheme may provide a more effective strategy for far-field thermal imaging. Inversely, when a point heat-source is placed near the outer boundary of the SSTM unit, the thermal energy is harvested into the center with unprecedented efficiency, as shown in Fig. 1 (c). This example only represents a small part of the potential of an SSTM unit, and more interesting functionalities that could be achieved through the combination of SSTM units will be demonstrated in the following.

We first experimentally demonstrate how to form a uniform heating region between four distant heat-sources by separating and enclosing them with four of our SSTM units. The fabricated structure is schematically illustrated in Fig. 2(a). It is noted that we have thermally isolated all of the fabricated samples by using an approximately $100\ \mu m$ thick layer of PDMS. The benefits of the PDMS layer are twofold -- heat convection by air is significantly reduced, and the PDMS layer on top of the sample is nearly "black" (i.e., 100% emissivity) for the wavelengths detected by the infra-red thermal imaging camera. The temperature distributions of four distant heat-sources with and without the SSTM units are quantitatively calculated in Fig. 2(b), in which the distance between two adjacent heat-sources (from the center of the left heat-source to the center of the right heat-source) is 13 cm. We can see that the valley-shaped temperature distribution for the case without the SSTM units become uniform after incorporating the four SSTM units. In the experimental setup, we use four brass cylinders with radius of 1 cm as heat-sources and the four boundaries are connected to a tank filled with ice water ($0\ ^\circ C$). The four brass cylinders are connected to a hotplate fixed at 60°. The cross-sectional temperature profile is captured with a Flir i60 infrared camera. The measured results with and without SSTM units are shown in Figs. 2(c) and 2(d), respectively, which is as expected and agrees very well with the theoretical results in Fig. 2(b). From the coordinate transformation perspective, this is because the geometrical size of heat-sources have been enlarged by $b/a$=5 times by using the SSTM. This property may find potential application in medical techniques like thermal therapy, where uniform heating is required over a body region.

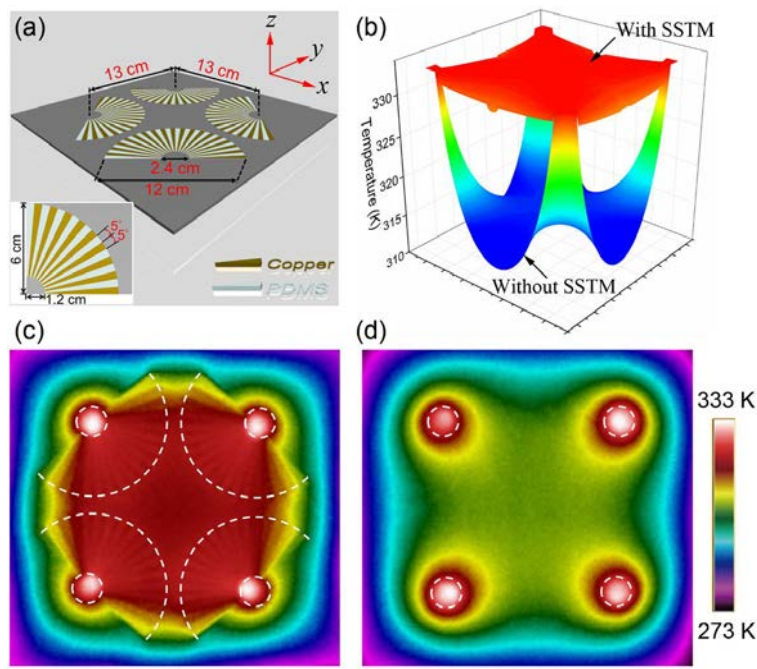

**FIG. 2** Experimental demonstration of forming a uniform heating region between four distant heat-sources enclosed by SSTM units. (a) Schematic of the fabricated sample. (b) Quantitative contrast of four distant heat-sources with and without SSTM. (c) Measured thermal profile of four distant heat-sources separately enclosed by SSTM units. (d) Measured thermal profile of four distant heat-sources without SSTM.

A natural question is whether a uniform heating region can be formed by only using copper of the same size. To answer this question, we simulated the case in which the SSTM units are replaced by four copper quadrants of the same size as that of Fig.2(a); see Fig. S1 in Supplementary Materials. It is clearly demonstrated that *sensu*-patterned copper exhibits higher temperature and better heating uniformity performance than the unpatterned bare copper counterparts of the same size overall. It is also important to examine the performance of SSTM cluster in Fig. 2(a) with larger distance between two adjacent units, which is demonstrated in Fig. S2 in Supplementary Materials. For quantitative comparison, the temperature of the center is marked out. In general, the performance is naturally degraded with an increase in *d*. In all individual cases of different separation distances, the temperature distribution for the case with SSTM is always much higher than the case without SSTM.

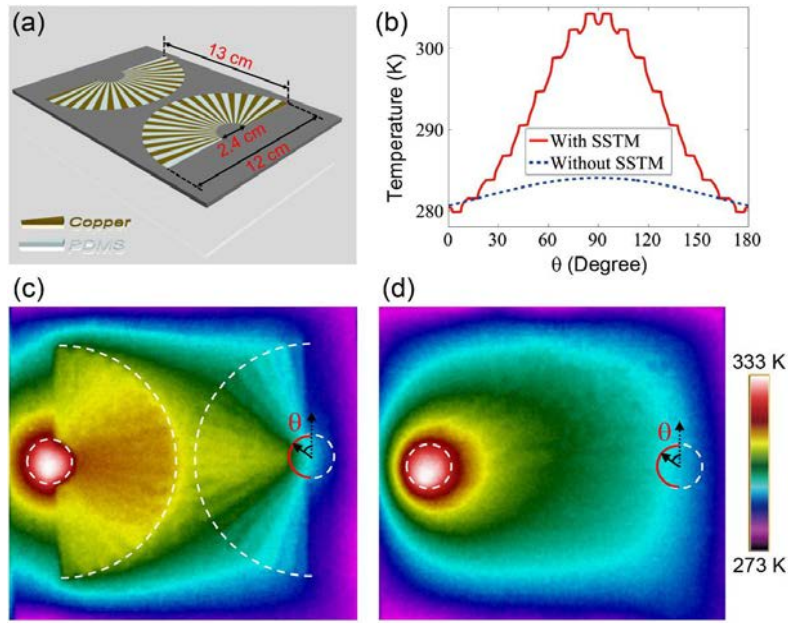

**FIG. 3** Experimental demonstration of thermal focusing. (a) Schematic of the fabricated sample. (b) Quantitative contrast of temperature distribution along semicircular red line in (c) and (d). (c) Measurement result of thermal focusing by enclosing a heat-source with SSTM unit and placing a second SSTM unit beside the first SSTM. (d) Measurement result of (c) without SSTM.

We next experimentally demonstrate heat flux focusing by using SSTM units. Consider a heat-source (brass cylinder) being enclosed by a SSTM unit, and a second SSTM is placed beside the first SSTM unit. We want to see that the enhanced thermal field of the heat-source is harvested by the second SSTM unit and focused into its center. The fabricated structure is schematically illustrated in Fig. 3(a). A brass cylinder with radius of 1 cm is placed at the center of the left SSTM unit and the four boundaries are connected to ice water (0 °C). The image captured by the infrared camera is shown in Fig. 3(c). For contrast, we measured the case where the two SSTM units are absent, as shown in Fig. 3(d). The temperature distribution along the semicircular red line in Figs. 3(c) and 3(d) is also calculated in Fig. 3(b), which agrees well with the experiment. Assuming the heat-source (placed in the first dotted circle) acts as a transmitter and the second dotted circle as a receiver, it is clear that the heat has been directed from the transmitter to the receiver by using the SSTM. When a thermal sensor is placed in the receiver position, the temperature is greatly enhanced, thus increasing the sensitivity of the temperature measurement. We also examine the cases where the SSTM units are replaced by copper wedges or copper quadrants (see Fig. S3 in Supplementary Materials). Obviously, there is no focusing effect with bare copper.

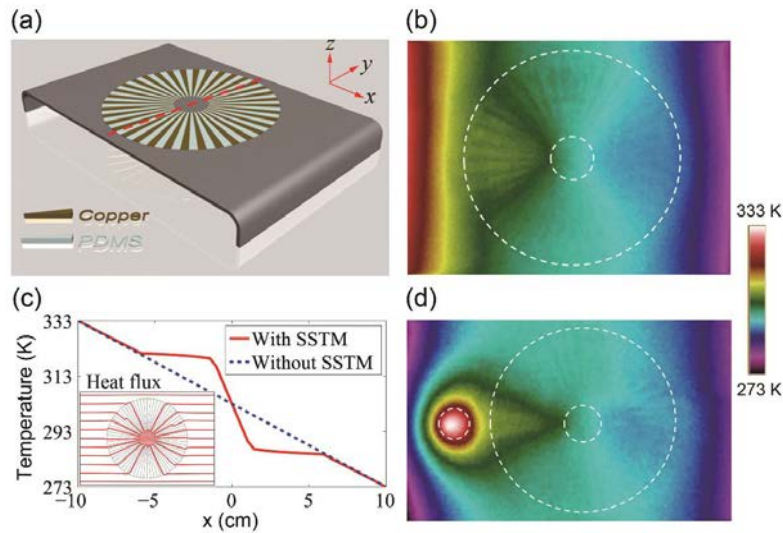

**FIG. 4** Experimental demonstration of an efficient thermal concentrator. (a) Schematic of the fabricated concentrator by the combination of two SSTM units. (b) Experimental verification of concentrating property in uniform thermal field. (c) Calculated temperature distribution along x-axis. The inset shows the heat flux lines. (d) Experimental verification of concentrating property in the presence of a point heat-source, emitting cylindrical heat fronts.

Moreover, we demonstrate an efficient heat flux concentrator by positioning two identical SSTM units. A concentrator is achieved by a combination of two back-to-back SSTM units, as shown in Fig. 4(a). For the measurement of the concentration effect, local heating on the left side is achieved by a hot plate fixed at $60\ °C$, and the right side is connected to a tank filled with ice water ($0\ ^\circ C$). The experimental result on heat flux concentration is presented in Fig. 4(b), which clearly shows the significant role of the SSTM's presence. Nearly all heat flux in the region ($0 \leq r \leq b$) is concentrated into the inner core ($0 \leq r \leq a$) without any reflection and distortion, indicating nearly perfect concentration. For quantitative comparison, the temperature distributions along *x*-axis are presented in Fig. 4(c) for both cases (i.e., with and without SSTM), which clearly reveals that the presence of SSTM (solid line) confines and builds up the temperature within the central area more efficiently than the case without the SSTM (dashed line). We also examine the concentration behavior in the presence of a point heat-source, as shown in Fig. 4(d). In the experimental setup of Fig. 4(d), the left and right boundaries are connected to ice water ($0\ ^\circ C$). Fig. 4(d) demonstrates the effectiveness of the proposed scheme in a non-uniform thermal field. In parallel, we study the cases in which the SSTM units are replaced by copper or PDMS, as shown in Fig. S4 in Supplementary Materials. Obviously, there is no concentration effect with bare copper or PDMS. Though thermal concentrators have been reported[17, 18], our scheme using SSTM units shows much higher concentration efficiency (nearly 100%). Beyond the thermal regime, it is noted that *dc* electric field concentrator based on resistor networks[30] and *dc* magnetic field concentrator based on superconductor- ferromagnetic metamaterials[31, 32] have been demonstrated recently.

The SSTM concept can be extended -- we find that the proposed SSTM approach can also be applied to manipulate *dc* currents, which has been numerically verified in Fig. S5 in Supplementary Materials. In the simulation setup, we use the same geometrical parameters and material components as those in the thermal regime. This means that the proposed SSTM is able to manipulate both thermal fields and *dc* currents simultaneously, demonstrating bifunctional property[33, 34]. In addition, all of the experimental setups in this paper are set as fixed boundary conditions ($0\ ^\circ C$). It is necessary to investigate the performance of the devices under convection boundary conditions. We take two heat-sources enclosed by SSTM units as example, and set all of the surfaces as convection boundary conditions with convection coefficient of 5 W/mK[18]. For continuous real-time observation, the sample with fixed temperature ($0\ ^\circ C$) or convection boundary conditions has been simulated with time evolution, as shown in the Supplementary

Materials, movies S1 and S2, respectively. From movie S2, we can see that a uniform heating region is formed at early times and disappeared at later times under convection boundary conditions. This is because the heat mainly conducts at early times and convects at later times.

In summary, we have proposed a new class of thermal metamaterials by using two regular bulk materials (copper and PDMS), and experimentally demonstrated its unique properties in terms of forming a uniform heating region, heat flux focusing, and concentration. These novel properties are robust to geometrical sizing of the SSTM unit whether scaled up or down. More interestingly the proposed SSTM is capable of manipulating both thermal field and *dc* currents simultaneously, exhibiting multi-functional property. Our scheme may find straightforward applications in technological devices such as thermoelectric devices, solar cells, thermal sensors, thermal imagers, as well as in thermal therapy applications. Our results can also provide novel ways to control other fields, e.g., dc magnetic fields, spin waves in spintronic devices, electrons in semiconductors. It has also introduced a new dimension to the emerging field of phononics: controlling and manipulating heat flow with phonons[7].

C.W.Q. acknowledges the Grant R-263-000-A23-232 administered by National University of Singapore. T.C.H. acknowledges the support from the National Science Foundation of China under Grant No. 11304253. T. Han, X. Bai, and D. Liu contributed equally to this work.

**References**

[1]   R. Venkatasubramanian, E. Siivola, T. Colpitts, B. O'Quinn, Nature **413**, 597 (2001).

[2]   B. C. Steele, A. Heinzel, Nature **414**, 345 (2001).

[3]   N. P. Padture, M. Gell, E. H. Jordan, Science **296**, 280 (2002).

[4]   P. Wang, S. M. Zakeeruddin, J. E. Moser, M. K. Nazeeruddin, T. Sekiguchi, M. Gratzel, Nat. Mater. **2**, 402 (2003).

[5]   B. Tian, X. Zheng, T. J. Kempa, Y. Fang, N. Yu, G. Yu, J. Huang, C. M. Lieber, Nature **449**, 885 (2007).

[6]   C. Chiritescu, D. G. Cahill, N. Nguyen, D. Johnson, A. Bodapati, P. Keblinski, P. Zschack, Science **315**, 351 (2007).

[7]   N. Li, J. Ren, L. Wang, G. Zhang, P. Hänggi, B. Li, Rev. Mod. Phys. **84**, 1045 (2012).

[8] M. Terraneo, M. Peyrard, G. Casati, Phys. Rev. Lett. **88**, 094302 (2002).

[9] B. Li, L. Wang, G. Casati, Phys. Rev. Lett. **93**, 184301 (2004).

[10] C. W. Chang, D. Okawa, A. Majumdar, A. Zettl, Science **314**, 1121 (2006).

[11] B. Li, L. Wang, G. Casati, Appl. Phys. Lett. **88**, 143501 (2006).

[12] H. S. P. Wong, S. Raoux, S. Kim, J. Liang, J. P. Reifenberg, B. Rajendran, M. Asheghi, K. E. Goodson, Proc. IEEE **98**, 2201 (2010).

[13] M. Maldovan, Phys. Rev. Lett. **110**, 025902 (2013).

[14] M. Maldovan, E. L. Thomas, Appl. Phys. Lett. **88**, 251907 (2006).

[15] S. Guenneau, C. Amra, D. Veynante, Opt. Express **20**, 8207 (2012).

[16] S. Guenneau, T. M. Puvirajesinghe, J. R. Soc. Interface **10**, 20130106 (2013).

[17] S. Narayana, Y. Sato, Phys. Rev. Lett. **108**, 214303 (2012).

[18] E. M. Dede, T. Nomura, P. Schmalenberg, J. Seung Lee, Appl. Phys. Lett. **103**, 063501 (2013).

[19] R. Schittny, M. Kadic, S. Guenneau, M. Wegener, Phys. Rev. Lett. **110**, 195901 (2013).

[20] S. Narayana, S. Savo, Y. Sato, Appl. Phys. Lett. **102**, 201904 (2013).

[21] T. Han, X. Bai, D. Gao, J. Thong, B. Li, C.-W. Qiu, Phys. Rev. Lett. **112**, 054302 (2014).

[22] H. Xu, X. Shi, F. Gao, H. Sun, B. Zhang, Phys. Rev. Lett. **112**, 054301 (2014).

[23] A. Alù, Physics **7**, 12 (2014).

[24] J. B. Pendry, D. Schurig, D. R. Smith, Science **312**, 1780 (2006).

[25] U. Leonhardt, Science **312**, 1777 (2006).

[26] Z. Jacob, L. V. Alekseyev, E. Narimanov, Opt. Express **14**, 8247 (2006).

[27] Z. Liu, H. Lee, Y. Xiong, C. Sun, X. Zhang, Science **315**, 1686 (2007).

[28] W. X. Jiang, C. W. Qiu, T. C. Han, Q. Cheng, H. F. Ma, S. Zhang, T. J. Cui, Adv. Mater. **25**, 6963 (2013).

[29] J. Li, L. Fok, X. Yin, G. Bartal, X. Zhang, Nat. Mater. **8**, 931 (2009).

[30] W. X. Jiang, C. Y. Luo, H. F. Ma, Z. L. Mei, T. J. Cui, Sci. Rep. **2**, 956 (2012).

[31] C. Navau, J. Prat-Camps, A. Sanchez, Phys. Rev. Lett. **109**, 263903 (2012).

[32] J. Prat-Camps, C. Navau, A. Sanchez, arXiv:1308.5878 2013.

[33] J. Y. Li, Y. Gao, J. P. Huang, J. Appl. Phys. **108**, 074504 (2010).

[34] M. Moccia, G. Castaldi, S. Savo, Y. Sato, V. Galdi, Phys. Rev. X **4**, 021025 (2014).

# Supplementary Information

## Manipulating Steady Heat Conduction by *Sensu*-shaped Thermal Metamaterials

Tiancheng Han[1*], Xue Bai[1,2,3*], Dan Liu[1,2,3*], Dongliang Gao[1], Baowen Li[2,3,4], John T. L. Thong[1,3], and Cheng-Wei Qiu[1,3]

[1]Department of Electrical and Computer Engineering, National University of Singapore, 4 Engineering Drive 3, Republic of Singapore.    *Equal Contribution

[2]Department of Physics and Centre for Computational Science and Engineering, National University of Singapore, Singapore 117546, Republic of Singapore

[3]NUS Graduate School for Integrative Sciences and Engineering, National University of Singapore, Kent Ridge 119620, Republic of Singapore.

[4]Center for Phononics and Thermal Energy Science, School of Physics Science and Engineering, Tongji University, 200092, Shanghai, China.

1. **Simulation for the creation of a uniform heating region by using SSTM or copper**

For quantitative comparison, the temperature of the center is marked out. We can see that the temperature of the center in Fig. S1(a) is higher than that in Fig. S1(b) 14.2 K. Clearly, the uniform heating performance of Fig. S1(a) (with SSTM) is much better than that in Fig. S1(b) (with bare copper).

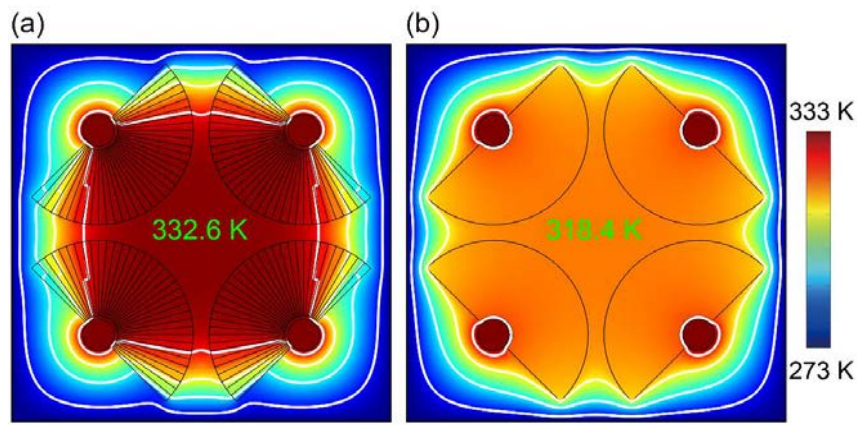

**Figure S1.** The temperature distributions of four distant point-sources respectively enclosed by SSTM or copper. (a) With SSTM. (b) With bare copper. For quantitative comparison, the temperature of the center is marked out. Isothermal lines are shown in white.

## 2. The performance of Fig. 2(a) with different distance between two adjacent SSTM units

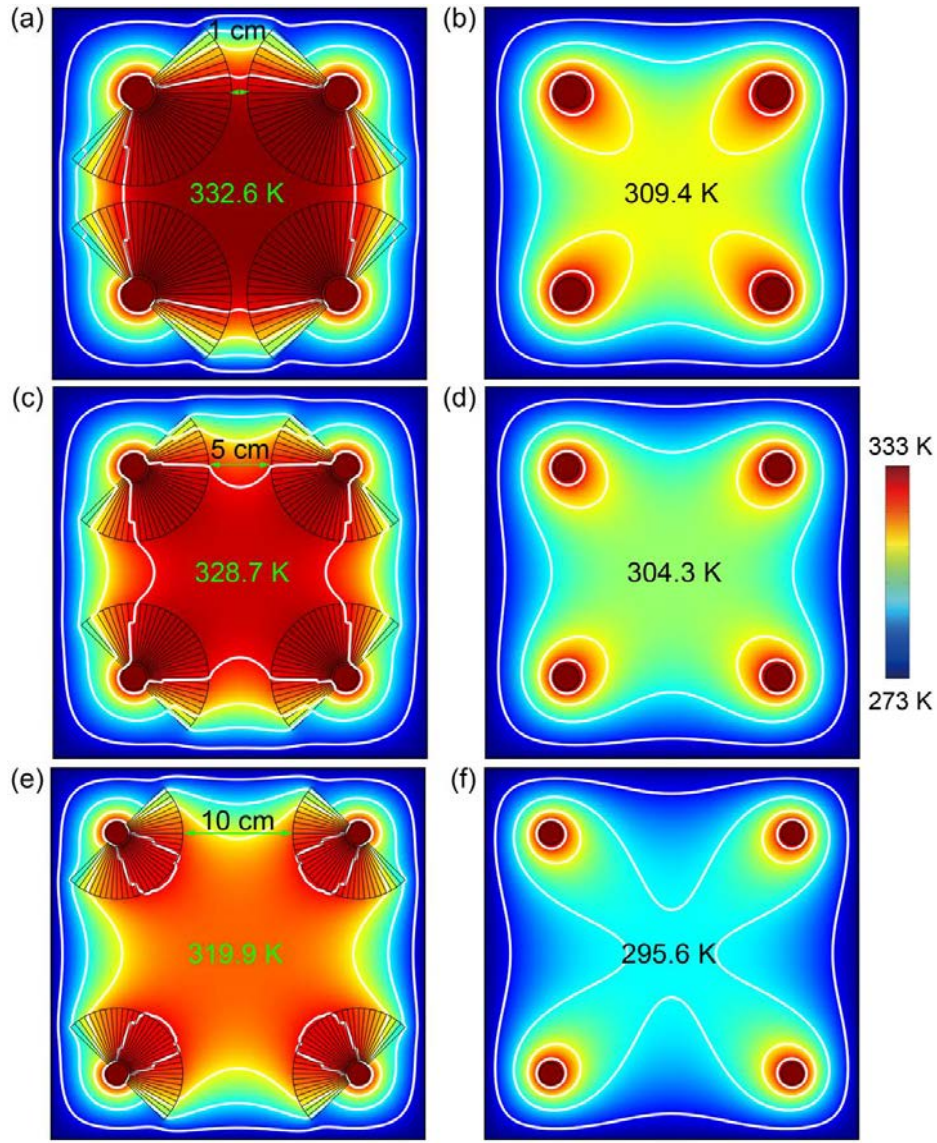

**Figure S2.** Performance of Fig. 2(a) with different distance between two adjacent SSTM units. Left column corresponds to simulated thermal profile of four distant heat-sources separately enclosed by four SSTM units when (a) $d$=1 cm, (c) $d$=5 cm, and (e) $d$=10 cm, respectively. Right column corresponds to simulated thermal profile of four distant heat-sources without SSTM when (b) $d$=1 cm, (d) $d$=5 cm, and (f) $d$=10 cm, respectively. For quantitative comparison, the temperature of the center is marked out. Isothermal lines are shown in white.

The performance is degraded with the increase of *d*. However, the temperature distribution for the case with SSTM (*d*=10 cm in Fig. S2(e)) is still much higher than the case without SSTM (*d*=1cm in Fig. S2(b)).

## 3. Thermal focusing effect with copper wedges, copper quadrants, or SSTM

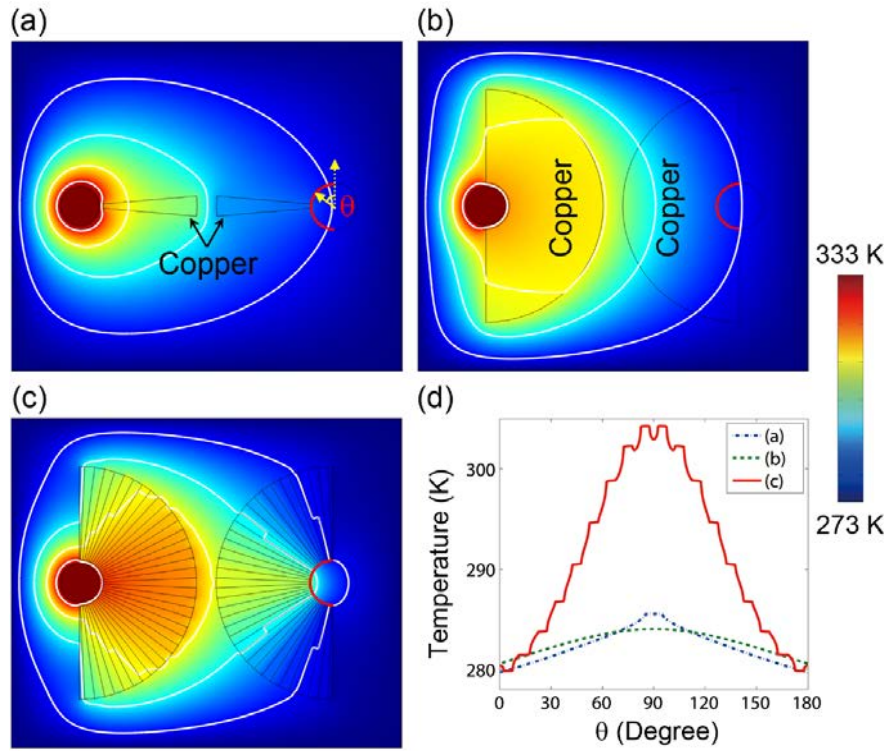

**Figure S3.** The thermal focusing effect with copper wedges, copper quadrants, or SSTM. (a) With copper wedges. (b) With copper quadrant. (c) With SSTM. (d) Quantitative contrast of temperature distribution along semicircular red line. Isothermal lines are shown in white.

## 4. Investigation of concentration effect with bare copper or PDMS

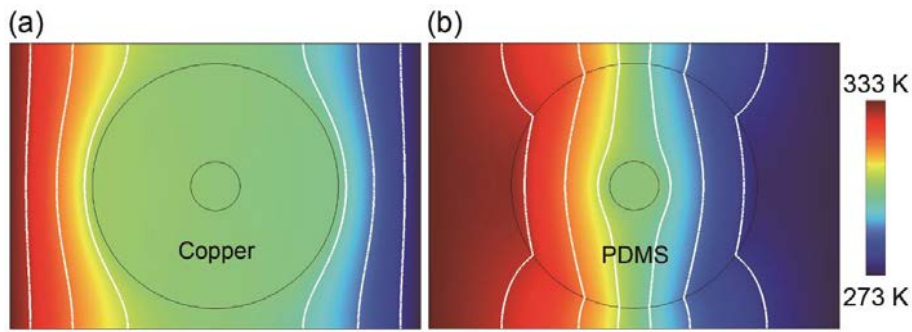

**Figure S4.** No concentration effect with bare copper or PDMS. (a) With bare copper. (b) With bare PDMS. Isothermal lines are shown in white.

**5. Simulation results of manipulating *dc* currents by using SSTM**

In the simulation setup, we use the same geometrical parameters and material components as those in the thermal regime. We choose *a*=1.2 cm and *b*=6 cm. An SSTM unit is composed of 18 copper wedges and 18 PDMS wedges, which is embedded in stainless steel. The electrical conductivities of copper and stainless steel are $\sigma_{Cu}=5.9\times 10^7$ S/m and $\sigma_b=1.3\times 10^6$ S/m, respectively. PDMS can be regarded as ideal insulating material. Figure S5 shows the simulation results of manipulating *dc* currents by using SSTM, which demonstrates bifunctional property of the proposed SSTM.

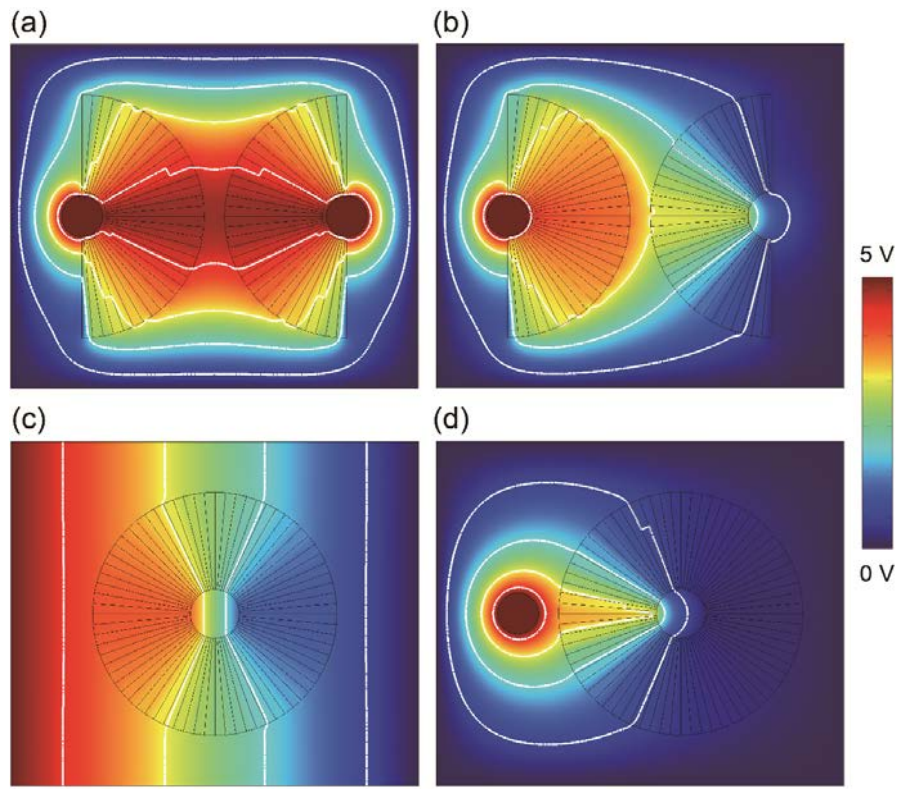

**Figure S5.** Simulation results of manipulating *dc* currents by using SSTM. (a) Forming a nearly constant voltage region between two cylindrical *dc*-sources. (b) Demonstration of *dc* focusing. Demonstration of *dc* concentration in uniform *dc* field (c) and non-uniform *dc* field (d). Equipotential lines are shown in white.